\documentclass[twocolumn]{autart}
\usepackage{natbib}
\usepackage{amssymb}
\usepackage{epsfig}
\usepackage{amsmath}
\usepackage{amssymb}
\usepackage{graphicx}
\usepackage{subfigure}
\usepackage[dvipsnames,usenames]{color}
\newcommand{\w}{\rm w}
\begin{document}
\begin{frontmatter}
\title{Nonlinear filtering in target tracking using cooperative mobile
  sensors\thanksref{footnoteinfo}}

\thanks[footnoteinfo]{This paper was not presented at any IFAC
meeting.}
\author[hu]{Jiangping Hu}\ead{hjp$\_$lzu@163.com},
\author[hu]{Xiaoming Hu\corauthref{cor}}\corauth[cor]{Corresponding author:  Xiaoming Hu. Tel. +46-8-790 7180.
Fax +46-8-22 53 20.}\ead{hu@math.kth.se}

\address[hu]{Optimization and Systems Theory and ACCESS Linnaeus Center, \\
                     Royal Institute of Technology, SE-100 44
                     Stockholm, Sweden}

\date{}

\begin{abstract}
Collaborative signal processing and sensor deployment  have been
among the most important research tasks in target tracking using
networked sensors. In this paper, the mathematical model is
formulated for single target tracking using mobile nonlinear scalar
range sensors. Then a sensor deployment strategy is proposed for the
mobile sensors and  a nonlinear convergent filter is built to
estimate the trajectory of the target.
\end{abstract}

\end{frontmatter}

\section{Introduction}
Sensor networks have the potential ability to monitor and instrument
the real world (\cite{agr,kum,feng,bro}). Among many challenges of
sensor networks, tracking target by using collaborative information
processing and optimal networking has been an important problem in
both theory and application.

In this paper, we  consider single target tracking by using a group
of mobile distributed nonlinear sensors such as range and direction
sensors in terms of  sensor deployment. If the measurement of the
target for each sensor is nonlinear and noisy, nonlinear filtering
theory has to be applied to estimate the state of the target. In
most cases, the target's dynamics is partially or even completely
unknown, which makes the modeling and estimation problem for target
tracking more complicated. Subsequently the target's dynamics will
be described by a linear Gaussian model and the distributed sensors'
measurements for the target will be nonlinear and noisy. Our aim is
to propose a convergent nonlinear filter to estimate the trajectory
of the target based on the measurements from the scalar range
sensors. Considering the sensors' tracking qualities,  sensor
deployment has to be dealt with by adopting formation control
techniques. This idea has also been adopted in, for example,
\cite{mart}, where the authors integrated motion coordination
strategies and extended Kalman filter to improve the tracking
performance of the sensor network. In \cite{cheng} collaborative
localization algorithms are proposed to reconstruct the position of
the target in a noise-free environment.

In many real applications, target tracking has to be dealt with in
the presence of disturbed target dynamics and nonlinear noisy
measurements. The estimation in nonlinear systems has been an
extremely important problem (\cite{juli}). Though there are now many
nonlinear filtering algorithms, such as point-based unscented
filters, density-based particle filters,  the extended Kalman filter
(EKF) has been a popular filter for nonlinear systems, in particular
for target tracking problems (\cite{mart,suga}). However, little
work has been performed to analyze the stability and convergence of
the filter, see \cite{reif} and references therein. It is well known
that the upper (lower) bound of the solution of the Riccati equation
associated with EKF depends on  the uniform controllability (uniform
detectability) of the considered nonlinear systems (see
\cite{baras}, Theorem 7). However, with the model used in this
paper, even though EKF has a comparative accuracy, it is difficult
to show the convergence of  EKF since the uniform complete
controllability cannot be assumed.  In addition, the Riccati
differential equations associated with the error covariance are very
difficult to integrate numerically.  Then an important problem
arises: can we design a convergent nonlinear filter to track the
target with unknown dynamics?

The contribution of this paper is twofold: one is the proposal of a
convergent nonlinear filter and the other is for appropriate
deployment of the sensors. Sensor deployment concerns the
connectivity and coverage of the network, which plays a key role in
energy conservation and monitoring quality (\cite{wang,kumar}) and,
at the same time, assures the feasibility of our proposed filter. A
fundamental problem, i.e., $k$-coverage problem, facing sensor
deployment for target tracking is how to deploy the sensors so that
every point in the target region will be monitored by at least $k$
sensors. For target tracking with mobile sensors, since the
deployment algorithm will be integrated with the estimation process,
$k$-coverage problems become more difficult to solve. In this paper,
inspired by a virtual vehicle approach in \cite{magnus},
neighbor-based formation control will be proposed to achieve an
equilateral-triangle deployment and solve the $3$-coverage problem
for mobile sensors.

The rest of the paper is organized as follows. In
Section~\ref{formulation}, we discuss the sensor deployment and
formulate the mathematical model in the target tracking problem. In
Section~\ref{filter}, on the basis of equilateral-triangle
deployment for mobile sensors, the design and convergence of a
nonlinear filter will be given to track the target. In
Section~\ref{motion},  the $3$-coverage problem will be solved for
mobile sensors by designing a decentralized control to achieve
equilateral-triangle formation.

Throughout this paper, we will use the following notations: $A$ is a
real matrix, $A^T$ denotes its matrix transpose, $tr(A)$ is the
trace of $A$, $\lambda_i(A)$ is an eigenvalue of $A$,  $\|A\|_2$ is
the spectral norm;  for vector $x$, $\|x\|_2$ is Euclidean norm and
$\|x\|_\infty$ is maximum norm; $<\cdot>$ denotes the inner product
in Euclidean space; $\otimes$ denotes the Kronecker product;
$E[\cdot]$ is the expectation operator; $col(\cdot)$ denotes the
concatenation; ${\rm Dirac}(\cdot)$ is the Dirac delta operator.

\section{Problem formulation}\label{formulation}

In this section, we will formulate the target tracking problem on
sensor deployment strategy and convergent nonlinear filter design.

Consider target tracking  with $N$ mobile sensors.  Suppose that
each sensor is modeled as two disks with radii $r$ and $r_c$, which
indicate the sensing range and the communication range,
respectively. In order to improve the target tracking quality,
having the target sensed by at least $k$ (no less than $3$)\;
sensors in a plane is critical. Meanwhile, the sensors should be
kept connected to transmit and receive data successfully. It has
been shown in \cite{wang} that if $r_c \geq 2r$, $k$-coverage
implies connectivity of the network. For the case the target region
is constrained in a space of dimension two, the problem of designing
a target tracking filter will be resolved in Section~\ref{filter}
and, the problem of how to make the region be $3$-covered by mobile
sensors in Section~\ref{motion}.


Let the kinematic equation of  sensor $i$ be described by a
nonlinear system
\begin{equation}
\label{senmot} \dot{s}_i(t)=f_i(s_i(t),u_{s,i}(t)),\; i=1, \cdots,
N,
\end{equation}
where $s_i(t)\in \mathbb{R}^n$ is the position of sensor $i$ and
$u_{s,i}(t)$ is its input. The measurement of the target for each
scalar sensor is nonlinear and noisy and can be expressed as
\begin{equation}
\label{meas} y_i(t)=\phi_i(p(t),s_i(t))+\omega_i(t),\; i=1, \cdots,
N,
\end{equation}
where  $\phi_i \in \mathbb{R}$ is continuously differentiable with
respect to $p$ and $\omega_i$ are independent Gaussian white noises
with covariance matrices $E[\omega_i(t_1)\omega_i^T(t_2)]=\Omega_i
{\rm Dirac}(t_1-t_2)$. The variable $p(t)\;(\in \mathbb{R}^{n_1})$
denotes the state of the target. When $p(t)$ represents the position
or orientation angle of the target, the sensors will be called range
sensors. Another kind of sensors is called velocity sensor if $p$
contains the velocity vector of the target.

Now suppose that the state $p(t)$ of the target evolves in a
continuous-time linear system with partially unknown input:
\begin{equation}
\label{targ}\begin{cases} \dot{p}(t)=Ap(t)+Bu(t),\\
p(t_0)=p_0,
\end{cases}
\end{equation}
where  $u \in \mathbb{R}^{n_2}\;(n_2\leq n_1)$ is the partially
unknown input, the initial state $p(t_0)$ is also an unknown
constant vector and $A, B$ are constant matrices. The unknown input
is generated by a linear exogenous system
\begin{equation}
\label{dist}\begin{cases}\dot{\zeta}(t)=\Gamma\zeta(t)+F\nu(t),\\
u(t)=D\zeta(t),
\end{cases}
\end{equation}
where $\zeta \in \mathbb{R}^{m}, \nu\in \mathbb{R}^{n_3}$, $\Gamma,
F, D$ are known matrices and the system disturbance $\nu$ is a
zero-mean Gaussian white noise with $E[\nu(t)\nu^T(s)]=\Psi{\rm
Dirac}(t-s)$.

Then the target's dynamics (\ref{targ}) and the kinematics
(\ref{senmot}) and the measurements (\ref{meas}) of $N$ sensors
compose an extended system
\begin{equation}
\label{exte1}
\begin{aligned}
\begin{pmatrix} \dot{p}\\\dot{\zeta}\end{pmatrix}
&=\bar{A}\begin{pmatrix}p\\\zeta\end{pmatrix}+\bar{B}\nu,
\end{aligned}
\end{equation}
\begin{equation}
\label{ext21}\dot{s}=f(s,u_s),
\end{equation}
\begin{equation}
\label{exte3} y=\phi(p,s)+\omega,
\end{equation}
 where $y=col(y_1,\cdots,y_N), s=col(s_1,
\cdots, s_N), u_s=col(u_{s,1}, \cdots, u_{s,N}),\;f=col(f_1, \cdots,
f_N), \omega=col(\omega_1,$ $\cdots, \omega_N)$ and
$$\bar{A}=\begin{pmatrix}A& BD\\ 0&
\Gamma\end{pmatrix},\;\bar{B}=\begin{pmatrix}0\\F \end{pmatrix}.$$

\begin{rem}
Here we cannot assume that system $(\bar A, \bar B)$ is (uniformly)
controllable. Thus, the boundedness of the solution of the Riccati
equation associated with EKF corresponding to system (\ref{exte1}),
(\ref{exte3}) can not be guaranteed.
\end{rem}

Let $C(p,s)$  be the Jacobian matrix of  $\phi(p, s)$ with respect
to $p$ and $\bar{C}=[C(p,s),\; 0_{N\times m}]$. If there is no
confusion, $C(p,s)$ will be abbreviated as $C$ in the sequel. When
the observability of time-varying system with $C$ or $\bar C$ is
mentioned, $p$ and $s$ are regarded as constants. In order to
investigate the observability of system
$\left(\bar{A},\bar{C}\right)$, we make an assumption as follows:

\begin{assum}
\label{obstra}The pairs $(A, C)$ and $(\Gamma, D)$ are observable
and no eigenvalue of  $\;\Gamma$ is a transmission zero of system
$(A, B, C)$.
\end{assum}

The following result can be checked by Hautus test and the proof is
omitted here (the detail can be found in \cite{hu}).
\begin{lem}
\label{obs} Under Assumption \ref{obstra}, the matrix pair
$\left(\bar{A},\bar{C}\right)$ is observable.
\end{lem}

\begin{rem}
For system $(\bar A, \bar B, \bar C)$ in this paper, we will show in
Section \ref{filter} that, in fact, $(\bar A, \bar C)$ will be also
uniformly detectable by finding a gain matrix $H(p,s)$ such that
$\bar A+H\bar C$ will have negative real-part eigenvalues for
arbitrary time-varying variables $p$ and $s$.
\end{rem}

In terms of Lemma \ref{obs}, a nonlinear observer-based filter can
be proposed for systems (\ref{exte1}) and (\ref{exte3}):
\begin{equation}
\label{obseq}\begin{cases} \dot{\hat{p}}=A\hat{p}+BD\hat{\zeta}+H_p(y-\phi(\hat{p},s)),\\
\dot{\hat{\zeta}}=\Gamma\hat{\zeta}+H_{\zeta}(y-\phi(\hat{p},s)).
\end{cases}
\end{equation}
In this paper, we limit ourselves to nonlinear observers that
satisfy the following constraints:  1) $H_p, H_{\zeta}$ are gain
matrices to be designed such that $H_pC(\hat{p},s)$ and
$H_{\zeta}C(\hat{p},s)$ are constant matrices; 2) the observer is an
asymptotic observer for the given system in the absence of system
disturbance $\nu$ and measurement noise $\omega$. The motivation for
the second constraint is that the observer should be an possible
asymptotic observer even when some, or all,  components of the
system state or the measurements are noise free.

The nonlinear function $\phi(p,s)$ can be expanded up to first order
via
\begin{equation}
\label{exp} \phi(p,s)=\phi(\hat{p},s)+C(\hat{p},s)(p-\hat{p})
+\Delta(p,\hat{p},s),
\end{equation}
where $\Delta(p,\hat{p},s)$ is the remaining nonlinear term. The
estimation errors are defined by $e_p=p-\hat{p},\;
e_{\zeta}=\zeta-\hat{\zeta}$. Then we can have the nonlinear error
dynamics
\begin{equation}
\label{obserr}
\begin{cases}
\dot{e_{p}}=(A-H_{p}C(\hat{p},s))e_{p}+BDe_{\zeta}-H_{p}\Delta(p,\hat{p},s)-H_{p}\omega,\\
\dot{e_{\zeta}}=\Gamma
e_{\zeta}-H_{\zeta}C(\hat{p},s)e_{p}-H_{\zeta}\Delta(p,\hat{p},s)
-H_{\zeta}\omega+F\nu.
\end{cases}
\end{equation}

To examine the error dynamics (\ref{obserr}),
consider a general Ito stochastic differential equation
\begin{equation}
 \label{ito} de(t)=g(e(t))dt+b(t)d\tilde{\omega}(t),
 \end{equation}
 where $\tilde{\omega}(t)$ is a standard Brown motion.

Then a differential generator associated with $e(t)$ is defined as
follows (\cite{zaka}):
\begin{equation}
 \mathcal{L}(\cdot)=\frac{\partial (\cdot)}{\partial e}g(e(t))+\frac{1}{2}tr(b(t)b^T(t){\rm Hess}(\cdot))
\end{equation}
where ${\rm Hess}(\cdot)$ denotes the Hessian matrix.

Now a key lemma to show the stochastic boundedness or stability is
given as follows (\cite{reif,zaka}):
\begin{lem}
\label{llyap} If there is a stochastic process $V(e(t))$ and
positive numbers $\lambda_1, \lambda_2, \lambda_3, \epsilon$ such
that
\begin{equation}
\lambda_1\|e(t)\|^2 \leq V(e(t)) \leq \lambda_2\|e(t)\|^2
\end{equation}
and
\begin{equation}
\mathcal{L}V(e(t)) \leq -\lambda_3V(e(t))+\epsilon
\end{equation}
are fulfilled, then the stochastic process $e(t)$ is exponentially
bounded in mean square, i.e.
\begin{equation}
E[\|e(t)\|^2] \leq
\frac{\lambda_2}{\lambda_1}\|e(0)\|e^{-\lambda_3t}
+\frac{\epsilon}{\lambda_1\lambda_3}
\end{equation}
for every $t \geq 0$.
\end{lem}

In the subsequent sections, we will firstly give a construction of
nonlinear filter (\ref{obseq}) and analyze its stability. Secondly
we will discuss a deployment for mobile sensors in order to ensure
the existence of nonlinear filter (\ref{obseq}) and improve the
tracking quality as well.

\section{Filter design and analysis}\label{filter}

In this section,  a nonlinear filter will be built to track the
target by using mobile range sensors.  Then the stochastic stability
of the filter will be analyzed.

\subsection{Target tracking filter design}

Suppose that the target dynamics (\ref{targ}) is  expressed as
follows:
\begin{equation}
\label{eq2-1}\begin{cases} \dot{x}=v,\\
\dot{v}=u,
\end{cases}
\end{equation}
where $x, v, u\in \mathbb{R}^n$ denote the position, velocity and
acceleration of the target, $p=col(x,v)$,
$$A=\begin{pmatrix}0&I_n\\0&0\end{pmatrix},\;
B=\left(\begin{array}{c}0\\I_n\end{array}\right).$$ Additionally,
the acceleration $u$ is given by the exogenous system (\ref{dist}).

In the rest of the paper, we will consider only scalar range sensors
whose measurements of target are given by
\begin{equation}
\label{eq2-3} y_i(t)=\phi_i(x(t),s_i(t))+\omega_i(t)\in
\mathbb{R},~i=1,\cdots,N
\end{equation}
or,
\begin{equation}
\label{eq2-4} y=\phi(x,s)+\omega\in \mathbb{R}^{N}.
\end{equation}

To reconstruct the state of the target, an assumption on the rank of
Jacobian matrix $C(x,s)$ is made:
\begin{assum}
\label{rank} For continuous differentiable function $\phi(x,s)$,
there exists a number $N$ such that the Jacobian matrix
$C(x,s)=\frac{\partial{\phi}}{\partial{x}}(x,s)$  is column full
rank.
\end{assum}

The assumption suggests that when enough number of sensors are
placed in a finite dimensional state space, we can always find that
the gradients $\frac{\partial{\phi_i(x,s)}}{\partial{x}}$ are
linearly independent and then the reconstruction of the states of
the target will be possible. As $N$ turns larger, the assumption on
Jacobian rank becomes satisfied more easily. For example, if we take
the measurement function as
\begin{equation}
\label{measexap} \phi(x,s)=col(\|x-s_1\|, \cdots, \|x-s_N\|).
\end{equation}
It follows that the Jacobian matrix is
$$col(\frac{(x-s_1)^T}{\|x-s_1\|},\cdots,
\frac{(x-s_N)^T}{\|x-s_N\|}).$$

When three or more sensors are placed in the plane, it is easy to
find non-collinear positions relative to the target for the sensors.
This is just the reason why we should consider the $3$-coverage
problem in mobile sensor network in next section.

In what follows, for the sake of constructing a  filter for the
extended system $(\ref{eq2-1}), (\ref{dist})$ and $(\ref{eq2-4})$,
we investigate the structure of  system $(\ref{dist})$. From
Assumption \ref{obstra},  $(\Gamma, D)$ is observable, then there
exists a matrix $G\in \mathbb{R}^{m\times n}$ such that $\Gamma-GD$
is a Hurwitz matrix. Then, for an arbitrary positive definite matrix
$Q_{\zeta}\in \mathbb{R}^{m\times m}$, there is a symmetric positive
definite matrix $P_{\zeta}\in \mathbb{R}^{m\times m}$ such that
\begin{equation}
\label{eq2-7}
P_{\zeta}(\Gamma-GD)+(\Gamma-GD)^TP_{\zeta}=-Q_{\zeta}.
\end{equation}

The following assumption is given to guarantee the convergence of
the proposed nonlinear filter (\ref{obseq}) associated with the
extended system $(\ref{eq2-1}), (\ref{dist})$ and $(\ref{eq2-4})$.

\begin{assum} \label{gassu}For gain matrix $G$ in equation
  (\ref{eq2-7}), we have the following two relations:
\begin{enumerate}
\item $P_{\zeta}(\Gamma-GD)G+\varpi D^TP_{1}=0$,
\item $P_1(DG-\frac{1}{\varpi}I_n)+(DG-\frac{1}{\varpi}I_n)^TP_1=-Q_1$,
\end{enumerate}
where $\varpi>1$ and $Q_1, P_1$ are $n$-dimensional symmetric
positive definite matrices.
\end{assum}

Now a nonlinear filter is proposed as follows:
\begin{equation}
\label{eq2-obs}
\begin{pmatrix} \dot{\hat{p}}\\\dot{\hat{\zeta}}\end{pmatrix}
=\begin{pmatrix}A& BD\\ 0&
\Gamma\end{pmatrix}\begin{pmatrix}\hat{p}\\\hat{\zeta}\end{pmatrix}+H(\hat{x},s)(y-\phi(\hat{x},s))
\end{equation}
where $\hat{p}=(\hat{x}^T,\hat{v}^T)^T, $ and the gain matrix
$H(\hat{x},s)= MJ_{\hat{x}}^{-1} C^T(\hat{x},s),$
\begin{equation*}
\begin{aligned}
M=&\begin{pmatrix}(\alpha-1)I_n-\frac{1}{\varpi-1}P^{-1}_{1}G^TD^TP_1\\
\frac{\alpha}{\varpi}I_n-\frac{1}{\varpi-1}P^{-1}_{1}G^TD^TP_1\\
G(\frac{\alpha}{\varpi}I_n-\frac{1}{\varpi-1}P^{-1}_{1}G^TD^TP_1)
-P_{\zeta}^{-1}D^TP_{1}\end{pmatrix},\\
J_{\hat{x}}=&\frac{\partial^T{\phi}}{\partial{x}}(\hat{x},s)
\frac{\partial{\phi}}{\partial{x}}(\hat{x},s)=\sum_{i=1}^{N}
\frac{\partial{\phi}_i}{\partial{x}}(\hat{x},s)
\frac{\partial^T{\phi}_i}{\partial{x}}(\hat{x},s),\\
 \alpha>&\frac{\varpi}{\varpi-1}.
\end{aligned}
\end{equation*}
With Assumption \ref{rank}, the inverse of $J_{\hat{x}}$ is well
defined.

\subsection{Stability analysis}
Now we turn to the stability analysis of filter (\ref{eq2-obs}).

Define $e_p=p-\hat{p},\; e_{\zeta}=\zeta-\hat{\zeta},$ then  the
nonlinear error dynamics (\ref{obserr}) can be rewritten in a
compact form:
\begin{equation}
\label{eq2-8}\dot{e}=\hat{A}e-H(\hat{x},s)\Delta(x,\hat{x},s)+
\bar{B}\nu-H\omega,
\end{equation}
where
$$
e=\begin{pmatrix}e_p\\
e_{\zeta}\end{pmatrix},
\hat{A}=\bar{A}-H(\hat{x},s)\bar{C}(\hat{x},s),
\bar{B}=\begin{pmatrix}0_{2n\times m}\\
F\end{pmatrix}
$$
and  $\Delta(x,\hat{x},s)$ satisfies
$$
\phi(x, s) =
\phi(\hat{x},s)+C(\hat{x},s)(x-\hat{x})+\Delta(x,\hat{x},s).
$$

\begin{assum}
\label{tayb} There are positive numbers $\delta, \chi$ such that the
remaining term $\Delta(x,\hat{x},s)$ is bounded by
\begin{equation}
\label{taybd} \|\Delta(x,\hat{x},s)\|\leq \chi \|x-\hat{x}\|^2,
\end{equation}
for $\|x-\hat{x}\|\leq \delta$.
\end{assum}


\begin{assum}
\label{jac} There is a positive number $\gamma$ such that the
smallest eigenvalue of matrix $J_{\hat{x}}$ is satisfied for all
$t\geq 0$:
\begin{equation}
\label{jacbd} \lambda_{\rm min}(J_{\hat{x}})\geq \gamma.
\end{equation}
\end{assum}

\begin{rem}
\label{jacrem} If we can resolve the $3$-coverage problem in the
mobile sensor network, this assumption will be fulfilled since
$J_{\hat{x}}$ is bounded positive definite when $N$ sensors are
deployed in the interested area (compact subset of Euclidean space)
with equilateral-triangle formation, which will be considered in
Section \ref{motion}.
\end{rem}

\begin{defn}\label{defn1}
In equation (\ref{eq2-8}), if there exist positive numbers $a_1,
a_2, a_3$ such that
\begin{equation}
E[\|e(t)\|^2] \leq a_{1} \|e(0)\|e^{-a_{2}t}+a_{3}
\end{equation}
holds for every $t \geq 0$, we say that the nonlinear filter
(\ref{eq2-obs}) can track the target.
\end{defn}

Now a main result will be presented as follows:

\begin{thm} \label{thmsen}Consider a target with disturbed dynamics (\ref{eq2-1}),
(\ref{dist}) and $N$ mobile sensors with nonlinear measurements
(\ref{eq2-4}). Under Assumptions \ref{obstra}, \ref{gassu},
\ref{tayb} and \ref{jac},  the filter defined by (\ref{eq2-obs}) can
track the target.
\end{thm}

Proof: Firstly, for system (\ref{eq2-8}), we will show that $(\bar
A, \bar C)$ with given gain matrix $H(\hat{x}, s)$ will be
stabilizable under Assumption \ref{obstra}. A variable transform is
defined as follows:
\begin{equation}\label{trans}
\tilde{e}=Te,
\end{equation}
where
$$
T=\left(\begin{array}{ccc}I_n& 0 & 0\\
0& I_n & 0\\
0& -G& I_m \end{array}\right).
$$
Then the error dynamics (\ref{eq2-8}) is transformed to
\begin{equation}
\label{errt}
\dot{\tilde{e}}=\tilde{A}\tilde{e}-TH\Delta(x,\hat{x},s)+
T\bar{B}\nu-TH\omega,
\end{equation}
where $\tilde{A}=T\hat{A}T^{-1}.$

Define $ V(\tilde{e})=\tilde{e}^TP\tilde{e},$ where $P$ is a
symmetric matrix such that
\begin{equation}
\label{lya}P=\left(\begin{array}{ccc}P_1& -P_1 & 0\\
-P_1& \varpi P_1 & 0\\
0& 0& P_{\zeta} \end{array}\right)\;(\varpi>1),
\end{equation}
and $P_1, P_{\zeta}$ and $\varpi$ satisfy Assumption \ref{gassu}. It
is easy to see that $P$ is positive definite by \emph{Schur
Complement Formula}. Obviously, $V(\tilde{e})$ is bounded by
\begin{equation}
\label{genb}\lambda_{\rm min}(P)\tilde{e}^T\tilde{e}\leq
V(\tilde{e})\leq \lambda_{\rm max}(P)\tilde{e}^T\tilde{e},
\end{equation}
where $\lambda_{\rm min}(P)$ and $\lambda_{\rm max}(P)$ are the
smallest and largest eigenvalue of $P$, respectively.

For the given stochastic process $V(\tilde{e})$, the differential
generator  is given by
\begin{equation}
\label{gene}\begin{aligned} \mathcal{L}V(\tilde{e})=&\tilde{e}^T
(P\tilde{A}+\tilde{A}^TP)\tilde{e}-2\tilde{e}^T
PTH\Delta(x,\hat{x},s)\\
&+tr(PT\bar{B}\bar{B}^TT^T)+tr(PTHH^TT^T).
\end{aligned}
\end{equation}

Under Assumption \ref{gassu}, it is not difficult to obtain that
$$
\tilde{A}=\left(\begin{array}{ccc}
\frac{1}{\varpi-1}P_1^{-1}G^TD^TP_1-(\alpha-1)I_n&I_n&0\\
\frac{1}{\varpi-1}P_1^{-1}G^TD^TP_1-\frac{\alpha}{\varpi}I_n&DG&D\\
P_{\zeta}^{-1}D^TP_1&(\Gamma-GD)G&\Gamma-GD
\end{array}\right),
$$
thus we have
\begin{equation}
\label{lyares}
Q=:-(P\tilde{A}+\tilde{A}^TP)=Diag\{2(\alpha-\frac{\alpha}{\varpi}-1)P_1,\varpi
Q_1, Q_{\zeta}\},
\end{equation}
which is a positive definite matrix due to Assumption \ref{gassu}
and $\alpha>\frac{\varpi}{\varpi-1}$. Thus, the matrices $\tilde{A}$
and $\hat{A}$ will be Hurwitz stable.

Let $\lambda_{\rm min}(Q)$ be the smallest eigenvalue of $Q$. By
Assumptions \ref{tayb}, \ref{jac}, then the generator is
$$
\begin{aligned}
\mathcal{L}V(\tilde{e}) \leq &-(\lambda_{\rm
min}(Q)-2\chi\lambda_{\rm max}(P)\|T\|_2\|H\|_2\|\tilde{e}\|)
\tilde{e}^T\tilde{e}\\
&+\lambda_{\rm max}(P)\|T\|_2^2(tr(FF^T)+tr(HH^T)).
\end{aligned}
$$
Since $\|H\|_2^2 \leq tr(HH^T)\leq \|M\|_2^2tr(J_{\hat{x}}^{-1})
\leq n\frac{\|M\|_2^2}{\gamma}$, we have
$$
\begin{aligned}
\mathcal{L}V(\tilde{e}) \leq  &-(\lambda_{\rm
min}(Q)-2\chi\lambda_{\rm max}(P)
\sqrt{\frac{n}{\gamma}}\|M\|_2\|T\|_2\|\tilde{e}\|)
\tilde{e}^T\tilde{e}\\
&+\lambda_{\rm max}(P)\|T\|_2^2(tr(FF^T)+n\frac{\|M\|_2^2}{\gamma}).
\end{aligned}
$$
Defining
$$
\tilde{\delta}=\min(\|T\|_2\delta, \frac{\lambda_{\rm
min}(Q)}{4\chi\lambda_{\rm
max}(P)\sqrt{\frac{n}{\gamma}}\|T\|_2\|M\|_2}),
$$
$$
\epsilon=\lambda_{\rm
max}(P)\|T\|_2^2(tr(FF^T)+n\frac{\|M\|_2^2}{\gamma}),
$$
and using (\ref{genb}) one obtains $ \mathcal{L}V(\tilde{e})\leq
-\eta V(\tilde{e}) +\epsilon $ with $\eta=\frac{\lambda_{\rm
min}(Q)}{2\lambda_{\rm max}(P)}$ for $\|\tilde{e}\|\leq
\tilde{\delta}$. According to Lemma \ref{llyap},
\begin{equation}\label{estbd}
E[\|\tilde{e}(t)\|^2] \leq \frac{\lambda_{\rm max}(P)}{\lambda_{\rm
min}(P)}\|\tilde e(0)\|e^{-\eta t} +\frac{\epsilon}{\lambda_{\rm
min}(P)\eta},
\end{equation}
by Definition \ref{defn1} and the transformation (\ref{trans}), the
conclusion follows. \hfill\rule{4pt}{8pt}



\section{Motion coordination of sensors}\label{motion}

In this section, we discuss conditions for deploying the mobile
sensors that guarantee Assumption \ref{rank} and Assumption
\ref{jac}, which play an essential role in the proposed filter
design. We will design neighbor-based control so the mobile sensors
move in equilateral-triangle formations.


Suppose that the interconnection topology of the sensor network is a
tree. For range sensors, the tree interconnection topology means
that sensor $i$ can measure the position (i.e.  distance and bearing
angle with respect to sensor $i$) of sensor $i-1$ for $i=1,\cdots,
N$. Additionally, suppose that all sensor can get the states of the
estimated target, which can be regarded as a virtual sensor labeled
$0$.

A unicycle model is used to describe the kinematics of sensor $i$
for $i=1,\cdots, N$, i.e.
\begin{equation}
\label{unic}\begin{aligned}
\dot{x}_{s,i}&=v_{s,i}\cos \varphi_i,\\
\dot{y}_{s,i}&=v_{s,i}\sin \varphi_i,\\
\dot{\varphi}_{i}&=\w_{i},
\end{aligned}
\end{equation}
where $s_i=(x_{s,i}, y_{s,i})^T\in \mathbb{R}^2$ is the position,
$v_{s,i}$ is the translational velocity, $ \varphi_i$ is the
orientation angle  and ${\rm w}_{i}$ is the angular velocity. Here,
the control to be designed is $v_{s,i}$ and ${\rm w}_{i}$.

\begin{figure}[!htp]
  \centering
  \includegraphics[width=0.4\hsize]{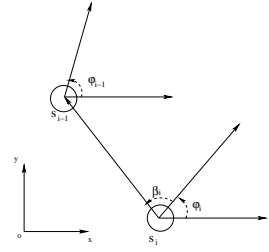}
  \caption{Notations of angles and distances}
  \label{unicfg}
\end{figure}

Let $d_i$ denote the actual distance between sensor $i$ and $i-1$
(while $d_{0,i}>0$ is the corresponding desired distance), let
$\beta_i$ be the actual bearing angle from the orientation of sensor
$i$ to edge $\overrightarrow{s_is}_{i-1}$ and $\beta_{0,i}$ the
corresponding desired bearing angle (see Fig. \ref{unicfg}). Here,
$\beta_{i}$ and  $\beta_{0,i}$ are constrained in the radian
interval $[-\pi, -\pi/2) \bigcup (-\pi/2, \pi/2) \bigcup (\pi/2,
\pi]$. Note that the desired formation is defined by $d_i=d_{0,i}$
and $\beta_{i}=\beta_{0,i}$ for $i=1, \cdots, N$. Finally, define
$\theta_i= \varphi_i- \varphi_{i-1}$.

Then we can rewrite the unicycle model (\ref{unic})  as
\begin{equation}
\label{unicmd}\begin{aligned}
\dot{d}_{i}&=-v_{s,i}\cos \beta_i-v_{s,i-1}\cos (\pi-\theta_i-\beta_i),\\
\dot{\theta}_{i}&=\w_{i}-{\rm w}_{i-1},\\
\dot{\beta}_{i}&=-{\w}_{i}+\frac{v_{s,i}}{d_i}\sin
\beta_i-\frac{v_{s,i-1}}{d_i}\sin(\pi-\theta_i-\beta_i).
\end{aligned}
\end{equation}

Inspired by the path-following control presented in \cite{magnus}, a
virtual sensor approach will be used. A reference point
$(x_{s,i}^0,y_{s,i}^0)$ is chosen on $s_i$'s axis of orientation at
a distance $d_{0,i}$ with bearing angle $\beta_{0,i}$. Then we have
\begin{equation}
\label{refpoi}\begin{aligned}
x_{s,i}^0&=x_{s,i}+d_{\rm 0,i}\cos (\varphi_i+\beta_{\rm 0,i}),\\
y_{s,i}^0&=y_{s,i}+d_{\rm 0,i}\sin (\varphi_i+\beta_{\rm 0,i}).
\end{aligned}
\end{equation}
Derivation of equation (\ref{refpoi}) in combination with unicycle
model (\ref{unic}) gives the following
 relationship:
 \begin{equation}
\label{refchag}
\begin{pmatrix}
v_{\rm s,i}\\\w_{i}
\end{pmatrix}=
\begin{pmatrix}\frac{\cos (\varphi_i+\beta_{\rm 0,i})}{\cos \beta_{\rm 0,i}}&
\frac{\sin (\varphi_i+\beta_{\rm 0,i})}{\cos \beta_{\rm 0,i}}\\
-\frac{\sin \varphi_i}{d_{\rm 0,i}\cos \beta_{\rm 0,i}}&\frac{\cos
\varphi_i}{d_{\rm 0,i}\cos \beta_{\rm 0,i}}
\end{pmatrix}
\begin{pmatrix}
\dot{x}_{s,i}^0\\\dot{y}_{s,i}^0
\end{pmatrix}-\begin{pmatrix}
0\\\dot{\beta}_{\rm 0,i}
\end{pmatrix}.
\end{equation}


On the one hand,  the initial values of $\beta_{0,i}$ can be taken
in $\beta_{0,i}\in [-\pi, -\pi/2)$ $\bigcup (-\pi/2, \pi/2)\bigcup
(\pi/2, \pi]$ in order to achieve equilateral-triangle formation
with tree interconnection topology. However, once the target changes
its orientation
 very fast, it is both necessary and
 efficient to decrease response formation time for target tracking
in practice. Hence, we need to regulate the desired bearing angles
timely.


An appropriate choice for the dynamics of the desired bearing angle
$\beta_{0,i}$ is given by:
 \begin{equation}
\label{bearag} \dot{\beta}_{0,i}=\left\{ \begin{array}{cl}
0&\mbox{if~}
 |\frac{\dot{\hat{x}}_0\ddot{\hat{y}}_0-\dot{\hat{y}}_0\ddot{\hat{x}}_0}{\dot{\hat{x}}_0^2+\dot{\hat{y}}_0^2}|\leq \varkappa,\\
-\frac{\varrho}{1+t^2}&\mbox{if~}
 \frac{\dot{\hat{x}}_0\ddot{\hat{y}}_0-\dot{\hat{y}}_0\ddot{\hat{x}}_0}{\dot{\hat{x}}_0^2+\dot{\hat{y}}_0^2}>\varkappa,\\
\frac{\varrho}{1+t^2}&\mbox{otherwise,}
\end{array}\right.
\end{equation}
for some threshold $\varkappa>0$ and $0\leq \varrho \leq 1$.

On the other hand, if $\dot{x}_{s,i}^0$ and $\dot{y}_{s,i}^0$  are
chosen as
 \begin{equation}
\label{refdyna0}\begin{aligned}
\dot{x}_{s,i}^0&=-k_i(x_{s,i}^0-x_{s,i-1})+\dot{x}_{s,i-1},\\
\dot{y}_{s,i}^0&=-k_i(y_{s,i}^0-y_{s,i-1})+\dot{y}_{s,i-1},
\end{aligned}
\end{equation}
for positive number $k_i$, the virtual sensor of sensor $i$ will be
driven to sensor $i-1$. Note that when $i=1$, $(x_{s,i-1},
y_{s,i-1})^T=(\hat{x}_{0}, \hat{y}_{0})^T$ will be the position of
virtual sensor $0$.

However, for range sensor $i \;(\geq 2)$, it is difficult  to obtain
the information of the velocity of its neighbor, i. e. sensor $i-1$,
so we need an assumption on the velocities of all mobile sensors:

\begin{assum}\label{velassu}
For both sensor $i \;(1\leq i\leq N)$ and the target, the velocities
are bounded, i.e., $\sqrt{\dot{x}_{s,i}^2+\dot{y}_{s,i}^2} \leq
k_s,$ and $\sqrt{\dot{x}_{0}^2+\dot{y}_{0}^2} \leq k_v,$ for some
positive $k_s$ and $k_v$.
\end{assum}

If the estimation error of the target's velocity is small enough,
from Assumption \ref{velassu}, the virtual sensor $i \;(1\leq i\leq
N)$ will still approach to sensor $i-1$ when the gains $k_i$ are
taken large enough and so we can ignore the velocity of sensor $i-1$
in equation (\ref{refdyna0}):
 \begin{equation}\label{refdyag1}
\begin{aligned}
\dot{x}_{s,i}^0&\cong-k_i(x_{s,i}^0-x_{s,i-1}),\\
\dot{y}_{s,i}^0&\cong-k_i(y_{s,i}^0-y_{s,i-1}).
\end{aligned}
\end{equation}


From the fact
$$
\begin{aligned}
x_{s,i}^0-x_{s,i-1}&=d_{0,i}\cos (\beta_{0,i}+\varphi_i)-d_i\cos (\beta_{i}+\varphi_i),\\
y_{s,i}^0-y_{s,i-1}&=d_{0,i} \sin (\beta_{0,i}+\varphi_i)-d_i\sin
(\beta_{i}+\varphi_i),
\end{aligned}
$$
and combining equations (\ref{bearag}) and (\ref{refdyag1}),  the
formation control (\ref{refchag})  for sensor $i$ is given by
 \begin{equation}
\label{fmcont}
\begin{aligned}
v_{\rm s,i}&=\frac{k_i}{\cos \beta_{0,i}}(d_i\cos (\beta_i-\beta_{0,i})-d_{0,i}),\\
\w_{i}       &=\frac{k_i}{d_{0,i}\cos \beta_{0,i}}(d_i\sin
\beta_i-d_{0,i}\sin \beta_{0,i})-\dot{\beta}_{0,i}.
\end{aligned}
\end{equation}

Then the result about the equilateral-triangle sensor deployment can
be stated as follows:

\begin{prop}
\textcolor[rgb]{1.00,0.00,0.00}{Under assumption that the velocity
of sensor $i-1$ is zero, for sensor $i$ }with the modified unicycle
model (\ref{unicmd}) with control (\ref{fmcont}) and any initial
values $\beta_{i}(0), \beta_{0,i}(0) \; (\in [-\pi, -\pi/2)\bigcup
(-\pi/2, \pi/2)$ $\bigcup (\pi/2, \pi])$, for all positive constants
$\epsilon_1, \epsilon_2$, there exist positive numbers $k_i,
\varrho$ and $T_0$ such that for every $t>T_0$
$$|d_i-d_{i,0}|<\epsilon_1,| \beta_i-\beta_{i,0}|<\epsilon_2.$$
\end{prop}

Proof:  Consider system (\ref{refdyag1}). Let
$e_x^i=x_{s,i}^0-x_{s,i-1},~e_y^i=y_{s,i}^0-y_{s,i-1}$, then we have
 \begin{equation}\label{error}
\begin{aligned}
\dot e_x^i&=-k_i e_x^i-\dot x_{s,i-1} ,\\
\dot{e}_y^i&=-k_i {e}_y^i-\dot y_{s,i-1} .
\end{aligned}
\end{equation}
Under  Assumption \ref{velassu} we have from \eqref{error} that
\begin{equation}
\begin{aligned}
\vert e_x^i(t)\vert &\le e^{-k_it}\vert e_x^i(0)\vert+max(k_v,k_s)/k_i\\
\vert e_y^i(t)\vert &\le e^{-k_it}\vert
e_y^i(0)\vert+max(k_v,k_s)/k_i.
\end{aligned}
\end{equation}
The conclusions follow since
$$d_{i,0}-\sqrt{e_x^i(t)^2+e_y^i(t)^2}\le d_i\le d_{i,0}
+\sqrt{e_x^i(t)^2+e_y^i(t)^2}$$ and
$$\tan(\phi_i+\beta_i)=\frac{d_0\sin(\phi_i+\beta_{i,0})-
  e_y^i(t)}{d_0\cos(\phi_i+\beta_{i,0})- e_x^i(t)}.\qquad\rule{4pt}{8pt}$$

\begin{rem}
When the sensors achieve the desired formation, Assumptions
\ref{rank} and  \ref{jac} will be guaranteed.
\end{rem}



\section{Conclusion}
In this paper, we designed a nonlinear observer-based filter to
track a single second order linear Gaussian target by using
measurements from $N$ scalar range sensors and analyzed the the
stability of the proposed filter in the sense of mean square. In
order to ensure the feasibility of the tracking filter, an
equilateral-triangle sensor deployment was proposed for the
convergence of the filter by neighbor-based formation control. A
very challenging topic for future research is the tracking of
multiple targets using mobile sensors. With issues such as data
association to overcome, it is far from trivial to extend the
results for a single target to such cases.

\bibliographystyle{harvard}

\end{document}